\documentclass[twocolumn,showpacs,preprintnumbers,amsmath,amssymb, superscriptaddress]{revtex4}

\usepackage{graphicx,subfigure}
\usepackage{dcolumn}
\usepackage{bm}
\usepackage{url}%

\begin{document}

\preprint{APS/123-QED}

\title{Search for Acoustic Signals from Ultra-High Energy Neutrinos\\ in 1500~km$^3$ of Sea Water}

\author{Naoko Kurahashi}
\affiliation{Departments of Physics and Applied Physics, Stanford University, Stanford, CA 94305}
 \email{naokok@stanford.edu}
\author{Justin Vandenbroucke}
\affiliation{Kavli Institute for Particle Astrophysics and Cosmology, Department of Physics and SLAC National Accelerator Laboratory, Stanford University, Stanford, CA 94305, USA}
\author{Giorgio Gratta}%
\affiliation{Departments of Physics and Applied Physics, Stanford University, Stanford, CA 94305}
\date{\today}

\begin{abstract}
An underwater acoustic sensor array spanning $\sim$1500~km$^3$ is used to search for cosmic-ray neutrinos of ultra-high energies (UHE, $E_{\nu}$$>$10$^{18}$~eV). Approximately 328 million triggers accumulated over an integrated 130 days of data taking are analysed. The sensitivity of the experiment is determined from a Monte Carlo simulation of the array using recorded noise conditions and expected waveforms. Two events are found to have properties compatible with showers in the energy range $10^{24}$~eV$<$$E_{sh}$$<$5$\times10^{24}$~eV and $10^{22}$~eV$<$$E_{sh}$$<$5$\times10^{22}$~eV. Since the understanding of impulsive backgrounds is limited, a flux upper limit is set providing the most sensitive limit on UHE neutrinos using the acoustic technique.  

\end{abstract}

\maketitle

\section{INTRODUCTION}

In the last few decades considerable attention has been devoted to understanding the origin, acceleration mechanisms, flux, and composition of the highest energy cosmic rays. Neutrinos are thought to play an important role at ultra-high energies (UHE, E$>$10$^{18}$~eV). Different mechanisms are being explored to probe the extremely small neutrino flux ($\ll$~1~km$^{-2}$~yr$^{-1}$) expected in this regime.  In the last decade, feasibility studies for acoustic UHE neutrino detectors have been initiated in large natural bodies of water, ice and salt.   The advantage to this technique is the possibility of building very large arrays (thousands of km$^3$) with sparse microphones, thanks to the large attenuation length of sound at the appropriate frequencies in these media. The Study of Acoustic Ultra-high Energy Neutrino Detection (SAUND) phase II is the first experiment to read out hydrophones undersea for the purpose of detecting UHE neutrinos using such an expansive array ($\sim$1500~km$^3$), and follows the first phase \cite{Vandenbroucke} where $\sim$15~km$^3$ were read out at the same site.

Acoustic radiation is produced from the volume expansion caused by the heating of the medium where the neutrino interacts and produces a shower.   Theoretical models~\cite{LearnedTheory} and experimental measurements \cite{Learned} show that the signature of this event is a single bipolar pulse, with acoustic energy concentrated around 10~kHz.
Noise conditions measured by SAUND II have been studied in detail~\cite{Kurahashi}. The analysis revealed that although the ambient noise correlates well with surface wind speeds, at higher frequencies the spectrum rolled off much faster than the expected power law due to the large depth of the hydrophones. The expected noise profile was parametrized and a good agreement was observed when compared to measured profiles.

Full data processing of pulse shapes, analysis of multi-hydrophone coincidences, and other techniques intended to reject background are reported here, along with a UHE neutrino flux limit set by a detailed simulation of the full experimental procedure.

\section{THE SAUND II EXPERIMENT}

\begin{figure}[htb!!!!!!!!!!!!!!!!!!!!!!!!!!!!!!!!!!!!]
\includegraphics[scale=0.5]{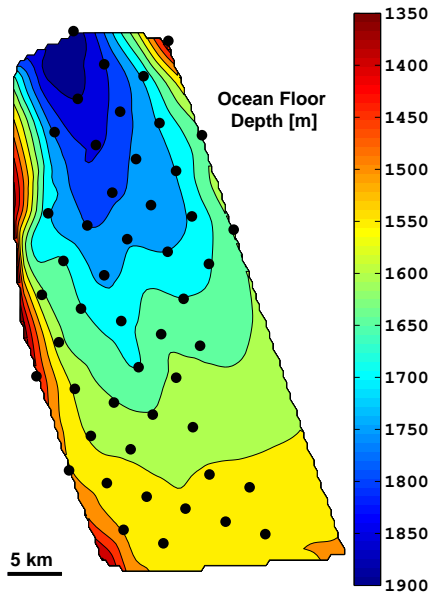}
\caption{Layout of the SAUND II array. Hydrophone locations are marked by bullets. Color shows the depth of the ocean floor.}
\label{Fig:ocean_floor}
\end{figure}

The second phase of SAUND (SAUND~II) employs a large hydrophone array at the US Navy's Atlantic Undersea Test and Evaluation Center (AUTEC) located at the Tongue of the Ocean (TOTO) in the Bahamas.    SAUND~II is currently the largest test for the feasibility of acoustic ultra-high-energy neutrino detection. This program follows a general study of the expected performance~\cite{Lehtinen}, and a first experimental phase (SAUND~I) using seven hydrophones ~\cite{Vandenbroucke}.  Since then, the US Navy has upgraded the hydrophones and the readout system of the array.  SAUND~II uses 49 of these hydrophones with digitized signals transmitted to shore over optical fibers.    The array spans an area of $\sim 20$~km $\times$ $50$~km with spacing of 3~to~5~km.   Hydrophones are mounted 5.2~m above the ocean floor, at depths between 1340 and 1880~m. Fig.~\ref{Fig:ocean_floor} shows the configuration and the topography of the ocean floor. 
Hydrophones are omnidirectional with a flat response (within 5~dB) at the frequencies considered. The gains of the 49 channels coincide to within 1~dB. Analog signals are regenerated at the shore station from the digital data (by the US Navy for their backwards compatibility) and fed to the SAUND~II data acquisition (DAQ) system that re-digitizes them at 156~kHz.    Since low frequencies are not relevant for SAUND~II, a hardware high-pass RC filter is applied to the analog data with a 3~dB point at 100~Hz.    

\begin{figure}[htb!!!!!!!!!!!!!!!!!!!!!!!!!!!!!!!!!!!!]
\includegraphics[scale=0.4]{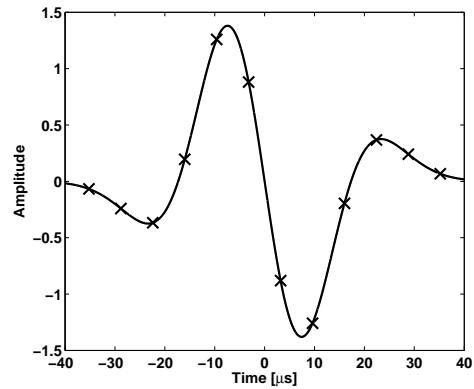}
\caption{Bipolar response function \cite{Vandenbroucke} used for online triggering. Markers show the digitized points used to match filter against the 156~kHz data sampling.}
\label{Fig:bipolar}
\end{figure}

A DAQ program with real-time analysis running on seven PCs records candidate neutrino events using a matched filter with the bipolar response function shown in Fig.~\ref{Fig:bipolar}. Because the phase response of the electronic system is unknown, SAUND II only considers the absolute value of the matched filter. Once per minute, the trigger threshold on each hydrophone is adjusted independently to a value that would have acquired 20~triggers/min in the previous minute. This differs from the SAUND I algorithm which only allowed thresholds to move up or down by a fixed step size. Each hydrophone triggers independently, and a 1~ms time series of only the triggered hydrophone is recorded.  In addition, every minute the threshold on each hydrophone is recorded. The noise conditions in the form of a power spectral density (PSD) and root mean square (RMS) are recorded every 5~s and 0.1~s respectively for each hydrophone.
The SAUND II DAQ system consists of seven National Instruments cards (NI-PCI-MIO-64)and eight PCs of which seven are dedicated to data acquisition and one for controlling the entire system. It also includes a front-end module with hardware high-pass filters and an IRIG timing signal distributor that connects to each of the cards. The NI ADC cards are controlled by COMEDI~\cite{comedi}, and the DAQ software uses KiNOKO~\cite{kinoko}.

\begin{figure}[htb!!!!!!!!!!!!!!!!!!!!!!!!!!!!!!!!!!!!]
\includegraphics[scale=0.4]{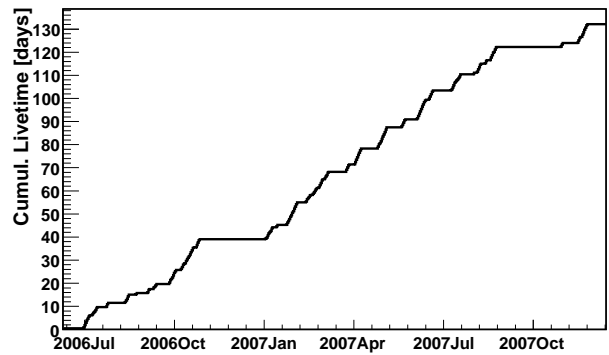}
\caption{Accumulated livetime of the SAUND II experiment.}
\label{Fig:livetime}
\end{figure}

By agreement with the US Navy, the SAUND II data acquisition system records data only when the array is not otherwise in use. From July 2006 to September 2007, the system ran under stable conditions for a total integrated time of more than 130~days as shown in Fig.~\ref{Fig:livetime}.

\begin{figure}[htb!!!!!!!!!!!!!!!!!!!!!!!!!!!!!!!!!!!!]
\includegraphics[scale=0.35]{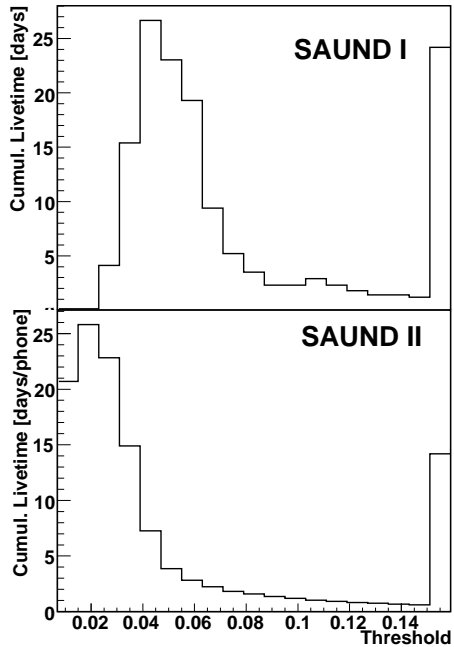}
\caption{Comparison of adaptive threshold levels used by SAUND I \cite{Vandenbroucke} and SAUND II. Threshold values have been normalized to account for the different sampling frequency and digitization of the response function (Fig.~\ref{Fig:bipolar}) in the two experiments. The quieter conditions in SAUND II are due to the absence of long coaxial cables, and an improved adaptive threshold algorithm.}
\label{Fig:thresholds}
\end{figure}

Thresholds used in  SAUND II are compared to those used in SAUND I in Fig.~\ref{Fig:thresholds}. Since the same region of sea is measured over several seasons, ambient noise in both experiments is expected to produce similar distributions. However, SAUND II measures lower levels of adaptive thresholds overall. This is due to a better electrical noise performance of the new array, owing to the absence of long coaxial cables and digital transmission of signals on fibers. 

\section{DATA ANALYSIS}
Table~\ref{Table} shows the total number of triggers accumulated and the size of the remaining data set after selection cuts are applied. Cuts 1 through 4 are applied on single-hydrophone triggers. After the source position in the ocean is triangulated using time difference of arrival, characteristics of these multi-hydrophone acoustic events are used on further cuts to test their consistency with expected properties of neutrino showers. 
In order to establish data analysis cuts that are minimally biased, an 8\% subset of data was initially used to establish the data reduction cuts. Because the noise environment can vary greatly depending on time of day and season, the subset was selected carefully to have a random distribution in time. Cuts 1 through 3 result from analysing the 8\% data alone. Once the full data was processed, it was found that some triggers are artificially generated by cross-talk between read-out channels. This effect was not discovered in the reduced data set as it affects a very small fraction of triggers. Therefore, parameters for cuts 4 to 7 have been modified on the full data set to eliminate triggers caused by cross-talk while maintaining efficiency for neutrino showers. The method used to validate this procedure is described in section~\ref{efficiency}.

In Table~\ref{Table}, the number of {\bf Online Triggers} is the total number of triggers recorded by the online DAQ system in all hydrophones. Minutes of individual hydrophones with high threshold ($>$0.04), high trigger rate ($>$500~triggers/min compared to the target 20~triggers/min set by the DAQ), high trigger rate on a nearest neighbor, or man-made noise in the water are excluded to select {\bf Quality Triggers}. Man-made noise can contaminate large parts of the array, and can be identified by narrow-band frequencies continuously dominating a wide area of hydrophones. This condition occurred a total integrated time of 9~min. Furthermore, data periods where the input cables appear to be disconnected are also removed. Such periods have unphysical, low noise levels, and are attributed to Navy personnel disconnecting individual cables for diagnosing problems in their system. One hydrophone had to be excluded entirely due to the majority of data belonging to this condition. For the remaining 48 hydrophones, this condition excludes $\sim$6\% of integrated livetime.  

\begin{table}[t]
\begin{center}
  \begin{tabular}{l c r}
         & Single-phone &       \\ 
    Cut  &   Triggers   & Events\\ 
    \hline
    1. Online Triggers      & 327.9M     &   --- \\ 
    2. Quality Triggers     & 146.7M     &   --- \\
    3. Waveform Selection  & 2,814,545  &   --- \\
    4. Single Phone Rate    & 2,562,047  &   --- \\  
    5. Triangulation        &     6,605  &  4,995\\
    6. Isolated Event       &     1,227  &    320\\
    7. Radiation Pattern    &         8  &      2\\
  \end{tabular}
\caption{Data reduction process. The remaining number of single-hydrophone triggers after each cut is shown in the second column. The third column shows the number of acoustic events which are formed by four triggers that triangulate to a point in the ocean.}
\label{Table}
\end{center}
\end{table}

\begin{figure}[htb!!!!!!!!!!!!!!!!!!!!!!!!!!!!!!!!!!!!]
\includegraphics[scale=0.45]{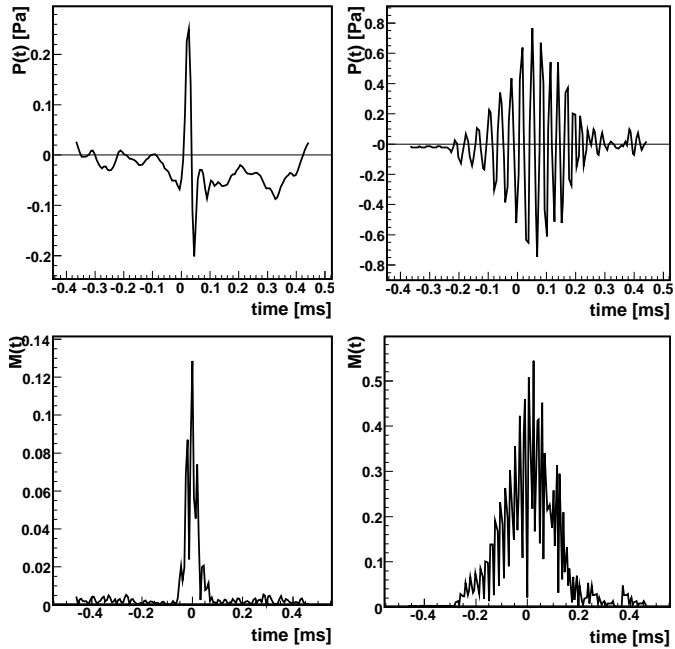}
\caption{Examples of Waveform Selection. The top panels show the time series, P(t), around two different triggers from the SAUND II data set. The panels at the bottom represent their matched filter time series M(t) calculated with the $\sim$76~$\mu$s bipolar response function, shown in Fig.~\ref{Fig:bipolar}, in 6.41~$\mu$s time steps. The GW parameter is obtained by smoothing M(t) and fitting to a Gaussian. The trigger on the left yields PAR=0.13 and GW=36.6~$\mu$s. The trigger on the right yields PAR=0.02 and GW=97.9~$\mu$s. Even though the trigger on the right has a higher matched-filter value at the time of trigger, the PAR and GW parameters correctly discriminate between a trigger consistent with a shower (left) and a background event (right).}
\label{Fig:waveshape}
\end{figure}

\begin{figure}[htb!!!!!!!!!!!!!!!!!!!!!!!!!!!!!!!!!!!!]
{
\includegraphics[scale=0.42]{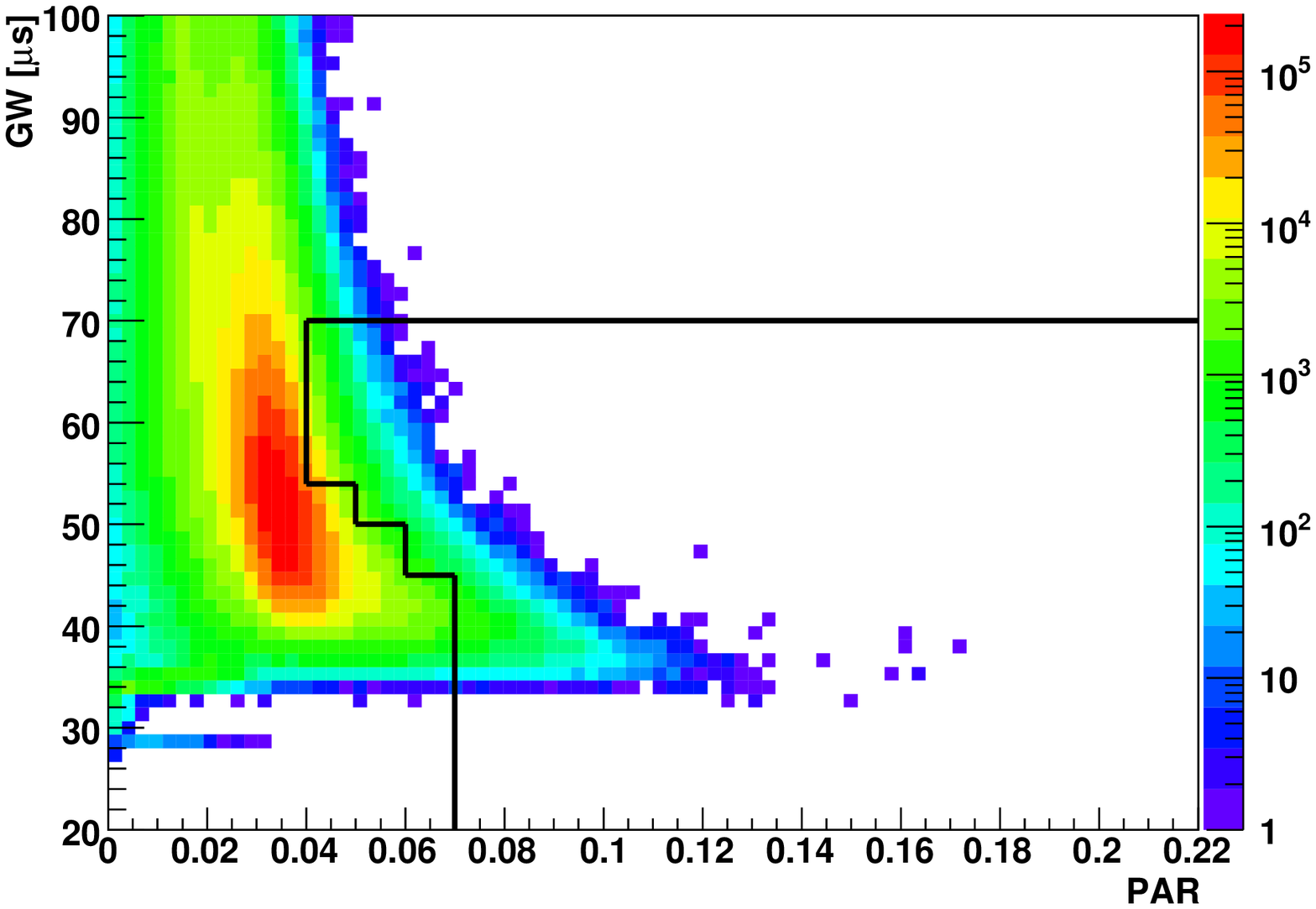}
}
{
\includegraphics[scale=0.42]{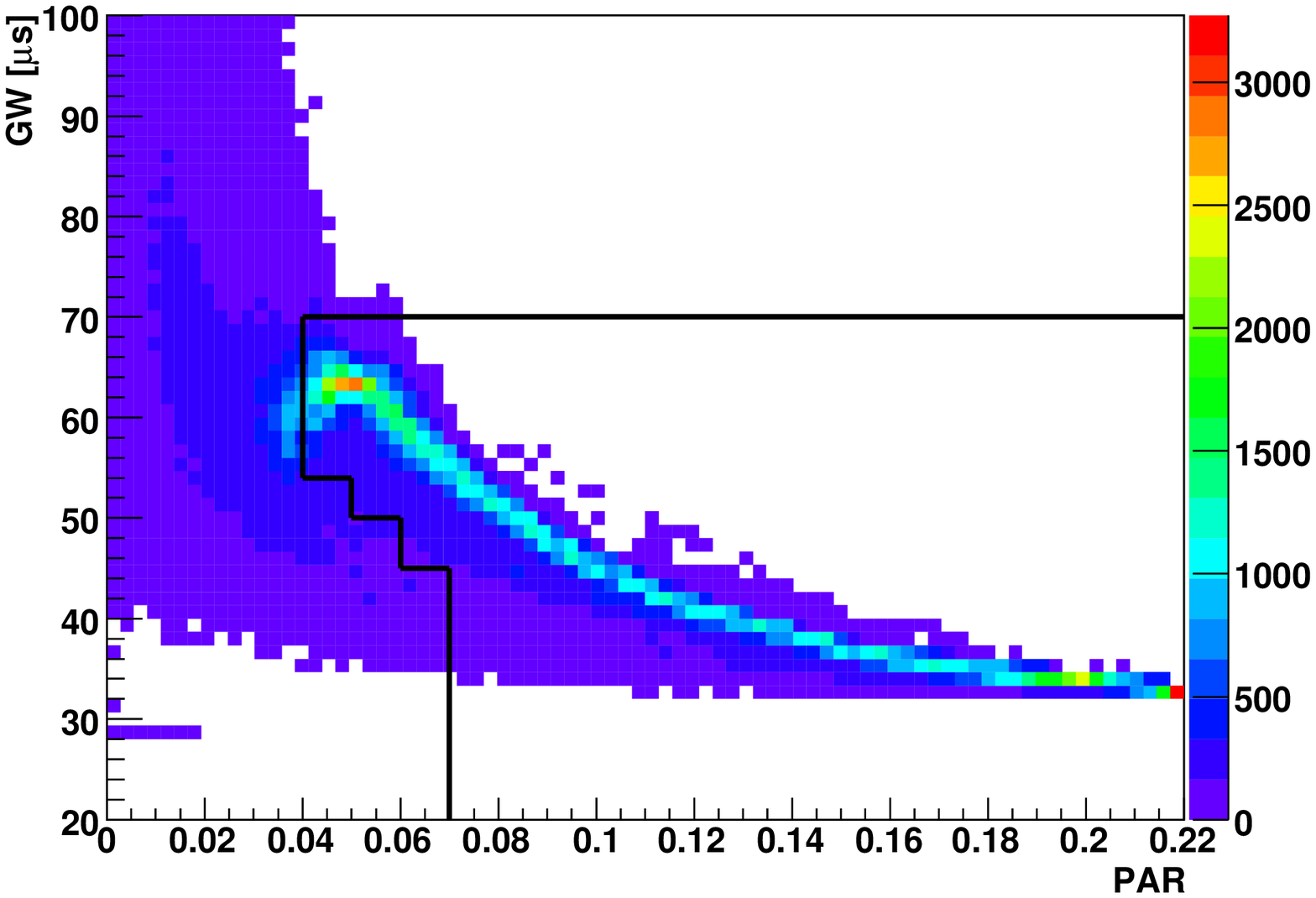}
}
\caption{Distribution of the parameters and accepted region of the Waveform Selection (top panel). The peak corresponds to the Knudsen noise floor~\cite{Kurahashi}. The 8\% subset of data used to establish this cut is plotted. The accepted region was optimized by generating over 300,000 Monte Carlo signals from 10$^{22}$--10$^{25}$~eV neutrino showers observed at angles -2$^\circ$--2$^\circ$ at various distances and adding recorded noise (bottom panel). }
\label{Fig:PAR}
\end{figure}

{\bf Waveform Selection} is used to further reduce background triggers. For each trigger, the online data acquisition system records a 1~ms waveform centered around the time of trigger. 
The absolute value of the matched-filter time series, $M(t)$, is calculated using the response function shown in Fig.~\ref{Fig:bipolar} for each time series. The Waveform Selection cut is based on two parameters from this time series: the peak area ratio (PAR) and the Gaussian width (GW). The PAR is obtained by dividing the value of $M$ obtained at the time of trigger by the integral of $M$ in the 1~ms window; PAR=$M(t_0)/\int_{t_0-0.5~ms}^{t_0+0.5~ms} M(t)dt$ where $t_0$ is the time of trigger. This quantity measures the significance of the trigger compared to its surrounding times. The GW is obtained by smoothing the time series of $M(t)$ to obtain the envelope shape, fitting this to a Gaussian, and extracting its width. For neutrino-induced signals, the width should be $\sim$40~$\mu$s. Many marine animals vocalize around frequencies of interest for this work. However, the signals originating from such sources tend to have a larger GW value. Fig.~\ref{Fig:waveshape} compares these parameters for two triggers from the data set. 
The distribution of these parameters and the cut region where neutrino showers are expected are shown in Fig.~\ref{Fig:PAR}.  
Triggers consisting of at least one data sample that saturate the ADC at $\pm$3.15~Pa bypass this cut and are kept regardless of the PAR and GW parameters. 

The {\bf Single Phone Rate} cut removes surviving triggers that occur on a single hydrophone clustered in time. Triggers accompanied by 10 or more triggers in any 5~s window are removed. In addition, triggers that occur less that 10~ms from each other are also removed. These rates are significantly higher than the 20~triggers/min target trigger rate set by the DAQ. Diffused, steady state UHE neutrinos are not expected to arrive and interact in bursts, so the removal of clustered triggers on single hydrophones only affects the livetime. Marine animals and artificial sources are usually responsible for such trigger patterns.
 
With the remaining triggers {\bf Triangulation} is performed. The 6,605 triggers produce 4,995 source locations in the ocean by requiring four hydrophones to have the appropriate difference in arrival times. The time difference of arrival method used is described in \cite{Vandenbroucke} and includes the effects of the depth-dependent speed of sound in the ocean. Only hydrophones that are within 8~km of the hydrophone of earliest arrival time are considered for each event. For each event, a local rectangular coordinate system tangent to the Earth surface, centered around the earliest arrival-time hydrophone is used. High rates of trigger on a single hydrophone in the coincidence time window ($\sim$8~s) produce combinatorics such that a single trigger can be involved in multiple events. However, by further excluding events that trigger the same hydrophone within a minute of each other, only 320 time {\bf Isolated Event}s survive. 

\begin{figure}[htb!!!!!!!!!!!!!!!!!!!!!!!!!!!!!!!!!!!!]
\includegraphics[scale=0.425]{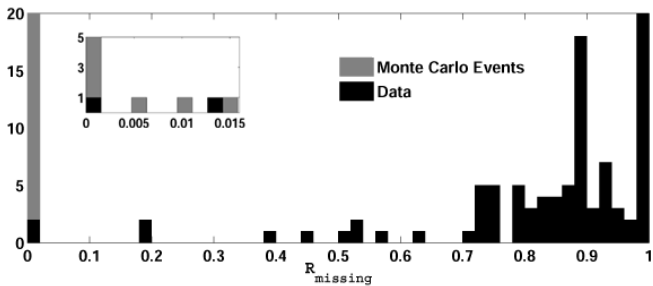}
\caption{Error metric $R_{missing}$. The overflow bin at $R_{missing}=1.0$ contains the 244 events with no possible shower geometries. The Monte Carlo events in the lowest bin extend beyond the maximum entries plotted to 131 events. The inset shows details of the distribution close to zero. Again, the number of Monte Carlo entries in the lowest bin extends to 128 entries. }
\label{Fig:metric}
\end{figure}

Since acoustic radiation from neutrino showers is not expected to be emitted spherically, but rather in a disk-like shape orthogonal to the direction of the shower~\cite{LearnedTheory, Lehtinen}, only showers of certain energy and orientation combinations can produce pulses capable of triggering all four hydrophones from the triangulated event location. This geometric constraint utilises the thresholds set at the time of the event, and only selects showers that are capable of triggering all four hydrophones with its acoustic radiation. However, not only does the radiation need to exceed the thresholds, but estimated pressure amplitudes must also be observed in the recorded waveforms. Only two events have observed pressures matching or exceeding those expected by geometrically possible showers, and pass the {\bf Radiation Pattern} cut. A metric, $R_{missing}$, quantifies the ratio of pressure missing in the recorded peak pressure, $P_{det}$, from the estimated peak amplitudes, $P_{est}$. 
\begin{equation}
R_{missing} = \frac{\sum_{i=1}^{4}(\Delta P)^2}{\sum_{i=1}^{4}(P_{est})^2}
\end{equation}
where
\begin{equation}
\Delta P = 
\begin{cases} P_{est}-P_{det} & \text{for $P_{est}>P_{det}$}
\\
 0 &\text{for $P_{est}<P_{det}$}
\end{cases}
\label{eq:pmetric}
\end{equation}
Since $\Delta P$ is defined as the pressure missing, it is set to zero for waveforms meeting or exceeding estimated pressure amplitudes.
$R_{missing}$ is calculated for a dense grid of shower energy-orientation combinations for each event. Fig.~\ref{Fig:metric} shows the minimum values obtained by considering all possible $E_{sh}$, $\theta$, and $\phi$ configurations for each of the 320 events. 244 events did not have possible showers with any direction and energy less than 5$\times$10$^{24}$~eV that fit the geometric constraint, and are assigned $R_{missing}=1.0$. 
Events with $R_{missing}<0.02$ pass the Radiation Pattern cut. The two events that pass require the shower energy to be $10^{24}$~eV$<$$E_{sh}$$<$5$\times$10$^{24}$~eV and $10^{22}$~eV$<$$E_{sh}$$<$5$\times10^{22}$~eV in order to be consistent with measured thresholds and peak pressures.

\section{EFFICIENCY AND SENSITIVITY ESTIMATES}
\label{efficiency}
The SAUND II detector is large enough to observe very different noise environments in different parts of the array. Therefore, the Quality Triggers cut described above removes livetime from single hydrophones minute-by-minute. Hence the configuration of enabled hydrophones is a complex pattern that varies from one minute to the next. To properly account for this, and to estimate the efficiency of the analysis cuts described above, neutrino events of different energies are generated by Monte Carlo simulation and injected randomly into the SAUND II livetime.   

The Monte Carlo study uses $\sim$66M simulated neutrino events generated with the following distribution in zenith angle $\theta$ and depth $z$, 
\begin{equation}
f_1(\theta,z) = Ae^{-z/lcos\theta} sin\theta
\end{equation}
where $l$ is the energy-dependent interaction length given by $l = 1/\sigma \rho N_A$. $\sigma$ is assumed to scale with energy as $\sigma\sim(E_{\nu}[GeV])^{0.363}$ \cite{Gandhi}, $\rho$ is the density of sea water, and $N_A$ is Avogadro's number.  The events are generated uniformly in the azimuthal angle $\phi$ as well as in latitude and longitude, in a 35~km$\times$57~km rectangular area completely encompassing the SAUND II array. The closest hydrophone-boundary distance is $>$3~km. The fraction of neutrino energy deposited into the hadron shower, $y$, is generated with the distribution
\begin{equation}
f_2(y) = d\sigma/dy_{total} = d\sigma/dy_{cc} + d\sigma/dy_{nc}
\end{equation}
where $d\sigma/dy_{cc} = (3+2(1-y)^2)y^{-0.67}$ and $d\sigma/dy_{nc} = (1+(1-y)^2)y^{-0.67}$ \cite{Beresnyak}. The hadron shower energy is obtained by $E_{sh}=y\times E_{\nu}$. We assume the same distributions for all flavors of neutrinos. Ray tracing is performed between each hydrophone-event pair including the effects of sound speed variation in the ocean to obtain arrival times.
The topography of the ocean floor is also considered. This produces some volume of ocean from where no rays can reach the hydrophones without intersecting with the ocean floor.  
An acoustic pulse is generated for each hydrophone-event pair using shower parameters~\cite{Alvarez-Muniz} and attenuation. For each event, a time of occurrence is assigned with a distribution that is uniformly random in the SAUND II livetime. This allows the assignment of realistic noise conditions from the data to each hydrophone in each event. 

Trigger candidates are then passed through the cuts in Table~\ref{Table}.  First the matched filter values of the Monte Carlo candidate triggers are compared to the threshold set at the occurrence time and hydrophone. The Quality Triggers cut keeps only the triggers that were assigned a timestamp with good run conditions, while the Waveform Selection cut discriminates on simulated pulse shapes with added noise.

The Single Phone Rate cut which is based on the rate of triggers on a single hydrophone can be affected by cross-talk. When a triggered waveform on one hydrophone produces cross-talk on other hydrophones that are also triggered, additional artificial triggers are created. The false increase in the trigger rate during this short period of time can affect the Single Phone Rate cut. To account for these situations, cross-talk triggers are generated by using the arrival times of surviving triggers and creating additional triggers on hydrophones within 8~km (electronic cross-talk is instantaneous compared to both the speed of sound in water and the sampling frequency of the ADC cards). The Single Phone Rate cut and Triangulation are then performed including these false triggers. None of the artificial cross-talk triggers were found to generate events. The parameters used for the Single Phone Rate cut were chosen to minimize the rejection of true events in the presence of cross-talk. 

The Isolated Event cut creates deadtime that is much less than 1\% of the integrated livetime. Finally Radiation Pattern fitting is performed to confirm that the metric $R_{missing}<0.02$, as shown in Fig.~\ref{Fig:metric}. 
 
\begin{figure}[htb!!!!!!!!!!!!!!!!!!!!!!!!!!!!!!!!!!!!] 
\includegraphics[scale=0.4]{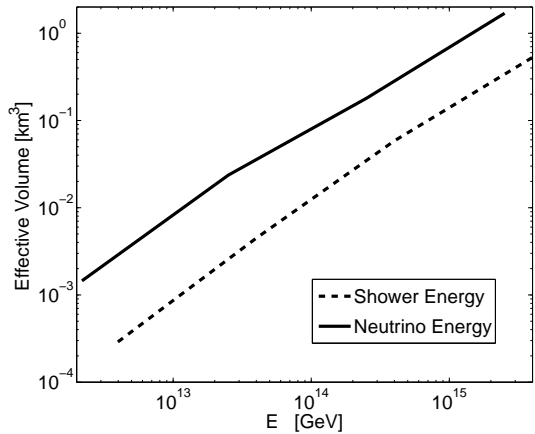} 
\caption{The effective volume of the SAUND II experiment for different neutrino and shower energies.} 
\label{Fig:efficiency} 
\end{figure}

\begin{figure}[htb!!!!!!!!!!!!!!!!!!!!!!!!!!!!!!!!!!!!]
\includegraphics[scale=0.425]{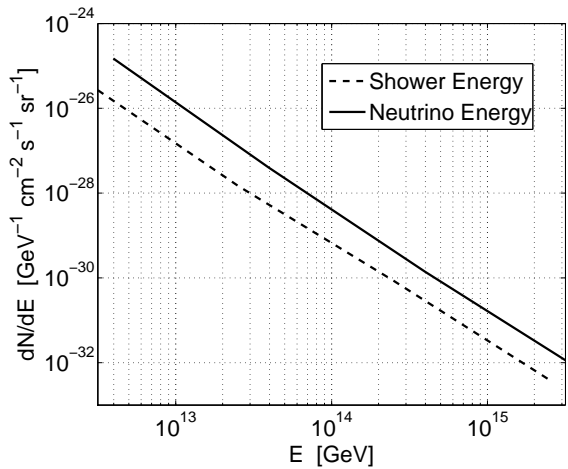}
\caption{Sensitivity to UHE neutrino flux at 90\% confidence level obtained by Monte Carlo study of the SAUND II experiment assuming no events are detected. The sensitivity is plotted for different neutrino and hadronic shower energies. Although on average neutrinos create hadronic showers that have $\sim$20\% of its energy, the two plots are not simply a 20\% shift from each other. This is due to the efficiency not scaling linearly with energy.}
\label{Fig:shower}
\end{figure}

The efficiency of the SAUND II detector, $\eta = {n_{detected}(E)}/{n_{generated}(E)}$, is obtained at each decade of neutrino energy and shower energy. The effective volume, $V_{eff} = V_{gen}\times\eta$, where $V_{gen}$ is the volume in which Monte Carlo events are generated, is shown in Fig.~\ref{Fig:efficiency}. Using the model-independent method described in~\cite{forte}, the sensitivity for UHE neutrino flux at different neutrino and hadron shower energies is plotted in Fig.~\ref{Fig:shower} using a 90\% confidence interval and assuming events remaining after cuts. 

\section{DISCUSSION AND RESULTS}
\begin{figure}[htb!!!!!!!!!!!!!!!!!!!!!!!!!!!!!!!!!!!!]
\includegraphics[scale=0.45]{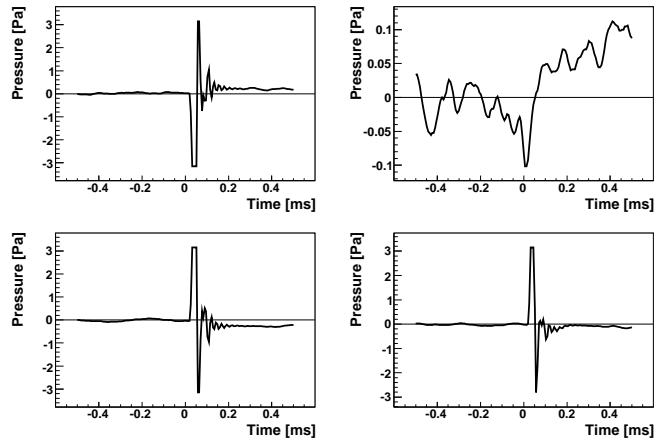}
\caption{Waveforms of the four triggers associated with an event occurring 29 September 2006, 22:36:01.62 UTC compatible with shower energy 5$\times$10$^{24}$~eV$>$$E_{sh}$$>$10$^{24}$~eV, zenith angle $12.0^\circ$$<$$\theta$$<$$16.0^\circ$, and azimuth $170.7^\circ$$<$$\phi$$<$$189.1^\circ$ north. Three of the triggers saturate the ADC at $\pm$3.15~Pa. Because the phase response of the electronic system is unknown, SAUND II only considers the absolute value of the matched filter, and the phase of the bipolar pulse is not used in this analysis.}
\label{Fig:wave2}
\end{figure}

\begin{figure}[htb!!!!!!!!!!!!!!!!!!!!!!!!!!!!!!!!!!!!]
\includegraphics[scale=0.45]{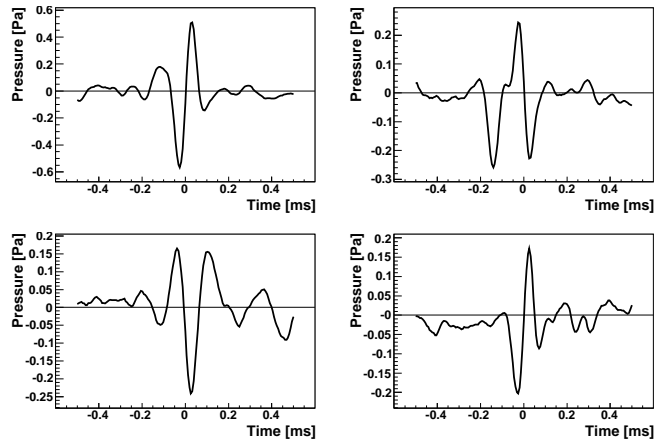}
\caption{Waveforms of the four triggers associated with an event occurring 3 May 2007, 15:01:04.17 UTC compatible with shower energy $10^{22}$~eV$<$$E_{sh}$$<$5$\times10^{22}$~eV, zenith angle $6.3^\circ$$<$$\theta$$<$$6.9^\circ$, and azimuth $233.8^\circ$$<$$\phi$$<$$235.5^\circ$ north.}
\label{Fig:wave3}
\end{figure}

In order to formalize the SAUND II results, the two events surviving all cuts must be accounted for. The four triggers associated with each event are shown in Fig.~\ref{Fig:wave2} and Fig.~\ref{Fig:wave3}. 
Further analysis capable of confirming the remaining two events as background or signal will require better understanding of the detector. Because the phase response of the system is unknown, the phase information of the triggered pulses is not utilised to discriminate events. Thus the event shown in Fig.~\ref{Fig:wave2} contains bipolar pulses with flipped phases, which are probably inconsistent as waveforms due to the same acoustic source. Apart from the phase, three of the four waveforms in Fig.~\ref{Fig:wave2} have similar high-amplitude bipolar pulses followed by a few more cycles of oscillation, while the fourth waveform on the top right has a much lower signal-to-noise. Since the probability of triangulating random arrival times to a single source location is extremely low, and since the four triggers are of order $\sim$100~ms to seconds apart, the origin of this event is assumed to be acoustic despite its peculiar features.  Furthermore, the three high-amplitude waveforms are similar to a type of background repeatedly seen in the SAUND II data set. All but the waveforms in Fig.~\ref{Fig:wave2} are eliminated due to their high repetition rate during their occasional bursts. The triggers in Fig.~\ref{Fig:wave2}, however, occur isolated in time and cannot be eliminated in such way.

As discussed in~\cite{Danaher}, the phase response of a hydrophone read-out system can distort bipolar pulses into more cycles, creating multi-polar oscillation signals. For this reason, the Waveform Selection cut does not define or utilize the number of cycles of oscillation. Thus some of the waveforms shown in Fig.~\ref{Fig:wave3} are not clearly bipolar. In addition, because many triggers occur with amplitudes that are not significantly higher than the noise level, it is difficult to define how many peaks constitute a signal. This effect also compounds the difficulty in phase matching bipolar pulses.

\begin{figure}[htb!!!!!!!!!!!!!!!!!!!!!!!!!!!!!!!!!!!!]
\includegraphics[scale=0.35]{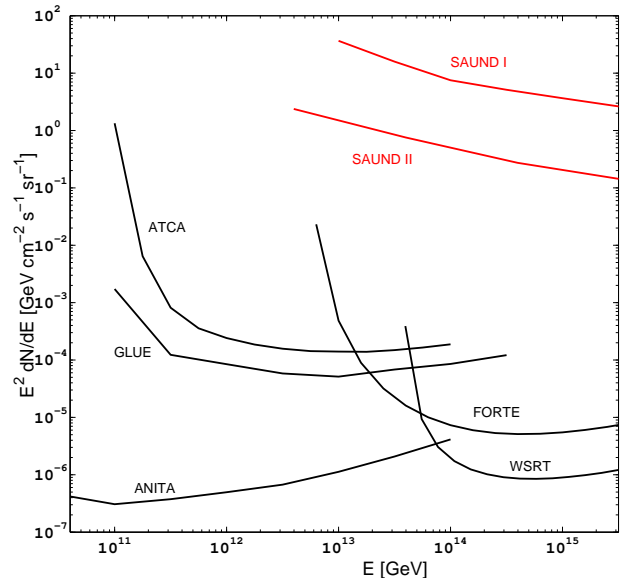}
\caption{Neutrino flux upper limit from the SAUND II experiment. Various other limits are plotted: SAUND I~\cite{Vandenbroucke}, GLUE~\cite{GLUE}, FORTE~\cite{forte}, ANITA~\cite{ANITA}, ATCA-LUNASKA~\cite{LUNASKA}, and NuMoon-WSRT~\cite{WSRT1, WSRT2}.}
\label{Fig:limits}
\end{figure}

Given the {\it a posteriori} observation made on two events, and the inability to perform a meaningful statistical analysis with such a small data set, an upper limit on the flux is set. Since the knowledge of impulsive backgrounds is incomplete, an upper limit with statistical significance on neutrino flux alone cannot be derived. Nevertheless, a conservative limit can be obtained by assuming that the two events are due to signal, and no other background is present. Following the treatment in~\cite{forte}, 
\begin{equation}
s = \int \lambda(\varepsilon)\Phi(\varepsilon)d\varepsilon
\end{equation}
where $s$ is the expected number of events, $\lambda$ is the sensitivity of the experiment at different energies, and $\Phi$ is the UHE neutrino flux. For a 90\% confidence level, 
\begin{equation} 
\int \lambda(\varepsilon)\Phi(\varepsilon)d\varepsilon \leqslant s_{up}
\end{equation}
where $s_{up}$ is the 90\% Poisson confidence interval upper limit of $n$ event detection. At each neutrino energy bin, $s_{up}(E_{sh})$ is calculated for different shower energy ranges $E_{sh}$. For shower energy ranges $10^{24}$--$10^{25}$~eV and $10^{22}$--$10^{23}$~eV, $n=1$ is used, making the upper bound $s_{up}=3.9$. For other shower energy ranges, $n=0$ is used, making the upper bound $s_{up}=2.3$. At each neutrino energy, a weighted average $\langle s_{up}\rangle$ is calculated using
\begin{equation}
\langle s_{up}(E_{\nu})\rangle=\sum_{E_{sh}}s_{up}(E_{sh})\frac{N(E_{sh})}{N(E_{\nu})}
\end{equation}
where $N(E_{sh})$ is the number of neutrinos that will produce a shower in the energy range $E_{sh}$ when $N(E_{\nu})$ neutrinos are generated at neutrino-energy range $E_{\nu}$. 
The weighted average gives $\langle s_{up}(E_{\nu})\rangle$ at each neutrino energy, and is used to set a flux limit
\begin{equation}
\Phi(E_{\nu})\lesssim\frac{\langle s_{up}(E_{\nu})\rangle}{E_{\nu}\lambda(E_{\nu})}.
\end{equation}
The limit obtained by this method along with other experimental limits are plotted in Fig.~\ref{Fig:limits}.

\section{CONCLUSION}
The main factors contributing to the difference in sensitivities of SAUND I and SAUND II are volume, threshold, and hydrophone spacing. While SAUND II is advantageous in the first two factors, its sensitivity is limited by the larger hydrophone spacing. In both arrays, hydrophones are mounted on the ocean floor, resulting in a mostly flat, two dimensional geometry which severely limits the sensitivity to the emission properties of showers, as already discussed in ~\cite{Vandenbroucke}.  

Unambiguously identifying rare UHE neutrinos from either acoustic or radio signals in huge naturally occurring bodies subject to varying and poorly understood backgrounds is very difficult.  Combining multiple techniques in a hybrid array featuring acoustic, radio, and/or optical detectors makes it possible to identify neutrinos by detecting two signals that have very different production and propagation mechanisms as well as backgrounds. One possible array has been simulated in~\cite{hybrid}. In the absence of such arrays, it is important to understand and identify backgrounds. The extensive study of the ambient noise condition using SAUND II data~\cite{Kurahashi} revealed that deep oceans are quieter than usually assumed at the high acoustic frequencies of interest. A model of such noise and a parametrization useful for ultra-high energy neutrino detection was developed. In executing a full analysis capable of detecting very rare events consistent with shower-induced signals, it was found that transient noise originating from discrete acoustic sources in the water causes the main limitation. A powerful cut, the Radiation Pattern cut, was developed to distinguish the geometry of emission to cope with such transient events. This cut will be significantly more effective when used with arrays of hydrophones arranged in all three dimensions. From such arrays, more information regarding the geometry of the emission lobe can be extracted.

\begin{acknowledgments}
We would like to thank the US Navy and, in particular, D.~Deveau and T.~Kelly-Bissonnette for their hospitality and help at the AUTEC site. We are grateful to D.~Kapolka (Naval Postgraduate School) and S.~ Waldman (Massachusetts Institute of Technology) for sharing their insights and expertise, and J.~Kerwin, K.~Montag, and S.~Wilson for their contributions through Summer Research College at Stanford University.  J. V. is supported by a Kavli Fellowship from the Kavli Foundation. This work was supported, in part, by NSF grant PHY-0457273.
\end{acknowledgments}

\bibliography{Kurahashi}

\end{document}